\documentclass[sigconf, nonacm]{acmart}
\AtBeginDocument{%
  }

\usepackage[T1]{fontenc}
\usepackage{babel}
\usepackage{hyphenat}
\usepackage{bbding}
\usepackage{pifont}
\usepackage{todonotes}

\usepackage{tikz}
\usetikzlibrary{shapes.symbols,arrows.meta,calc,backgrounds,fit,decorations.pathreplacing}
\usepackage{amsmath}
\usepackage{subcaption}
\usepackage{float}
\usepackage{hyperref}
\usepackage{graphicx}
\usepackage{xcolor}
\usepackage{tabularx}
\usepackage{booktabs}
\usepackage{multirow}
\usepackage{makecell}
\usepackage{wrapfig}

\usepackage{newunicodechar}

\definecolor{ocre}{RGB}{243, 102, 25}
\definecolor{graph_green}{HTML}{A3D977}
\definecolor{graph_orange}{HTML}{F8A17A}

\newcommand{\proxyname}[0]{{Stork}}
\definecolor{colgreen}{HTML}{82B366} 
\newcommand{\cmark}{\LARGE{\textcolor{graph_green}{\checkmark}}}
\newcommand{\xmarkr}{\LARGE{\textcolor{graph_orange}{$\times$}}}

\newcommand{\asia}[0]{{Japan }}
\newcommand{\NA}[0]{{US }}

\newcommand{\smartparagraph}[1]{\vspace{.05in}\noindent\textbf{#1}}

\begin{document}

\date{}

\title{Explicit QUIC Proxies for Server-Side Geo-blocking Bypass}



\author{Aurélien Buchet}

\author{Soyong Kim}

\author{Tom Barbette}

\author{Cristel Pelsser}

\email{{firstname}.{lastname}@uclouvain.be}
\affiliation{
   \institution{Université Catholique de Louvain}
    \city{Louvain-la-Neuve}
    \country{Belgium}
}

\begin{abstract}

Geo-restricted content is increasingly common on the Internet, forcing users to rely on circumvention techniques, such as VPNs, to access the web from a seemingly different location.
However, these often come with a financial cost and can degrade performance.
The rise in popularity of the QUIC protocol, which allows connections to migrate between paths, opens opportunities to circumvent such restrictions.
We scan web servers and find that a large portion of geo-blocked content is enforced on the server, at the application layer, rather than on-path. This check is performed once, when the request arrives, and is not repeated as the connection continues.
This allows a client to issue its request from a whitelisted IP address and, once the server has accepted it, migrate the connection to an otherwise unauthorized address for the rest of the transfer (post-header migration). 
It bypasses the block while maximizing direct traffic, thereby reducing eventual circumvention-related costs.
Building on this insight, we introduce \proxyname{}, an HTTP/2-to-HTTP/3 web proxy that bypasses geo-blocking while introducing negligible additional latency.
We demonstrate that our solution is compatible with popular clients and servers.
In controlled experiments with 2\,MB requests, our proxy migrates 99\% of the transferred data onto the unauthorized path. Across real-world targets that support QUIC migration, it migrates at least 75\% of the data for more than 52\% of them.
On real geo-blocked content, post-header migration bypasses the block for 90\% of domains, whereas post-handshake migration, as used by prior work, succeeds for only 60\%.
YouTube, one of the most widely used services on the Internet, is where the approach matters most: a video is not one transfer but a long series of chunk requests, each of which the server re-authorizes on arrival. Migration must therefore be repeated for every chunk, which post-handshake migration cannot do at all, while \proxyname{} still delivers more than 90\% of the content over the direct path for a set of 37 geo-restricted videos.

\end{abstract}

\sloppy

\maketitle

\section{Introduction}

Open access for content on the Internet is a fundamental principle of the web.
Yet, in practice, achieving it has become more and more difficult.
In addition to censorship coming from governments and ISPs that are widely studied~\cite{sundara2020censored,niaki2020iclab}, content providers also restrict access to some of their content at the application layer based mainly on the client IP address. An illustrative case is the websites that blocked access from European IPs when GDPR was introduced to avoid having to comply with it~\cite{bbc_gdpr_2018}.

Nowadays, the most popular ways to access blocked content rely on tunneling traffic through authorized networks: basic VPNs, censor-resistant VPNs such as Psiphon, or multi-hop, anonymity-preserving networks such as Tor~\cite{tor}.
However, these solutions often come with a performance cost due to needing to route traffic through a third party, which may be located far from the client or the server.
For instance, Xianghang et. al.~\cite{mi2019resident} study the rise of residential proxies (RESIP) using residential (sometime mobile) devices as relays to bypass server-side blocking.
Their appeal against \emph{server-side} blocking is that RESIP exit addresses belong to ordinary consumer ISP ranges. As they are hardly distinguishable from that of a legitimate customer, they are less likely to be blocked as opposed to datacenter-based VPNs that geo-blocking services list and reject wholesale. However, they come with limited bandwidth, total data caps, higher price and/or less reliability~\cite{khan2023stranger}.

Geo-blocking systems detect a client's location based on the source IP address and block the request if the client is located in a restricted region.
The tools implementing this are application-layer components: the IP2Location geo-blocking plugin for WordPress~\cite{ip2location} and Cloudflare's Web Application Firewall~\cite{cloudflareWAF} both inspect the HTTP headers and the source IP \emph{at the time of the request}.

In recent years, the adoption of the QUIC protocol~\cite{rfc9000}, designed as a replacement for TCP and TLS for web traffic, has been increasing rapidly~\cite{ruth2018first,Over9000,zirngibl2024quic,buchet2025analysis}.
QUIC enables connection migration and multi-pathing: the ability to switch between, or simultaneously use, different network interfaces (e.g., Wi-Fi and cellular) without interrupting the connection.
We observe that QUIC connection migration and multi-pathing techniques (MPTCP~\cite{rfc8684} and MPQUIC~\cite{de2017multipath}) therefore pose a vulnerability to geo-blocking websites: a client can migrate right after making the initial request and receive the content over a different path that would otherwise be blocked. Here we focus on QUIC, as our measured MPTCP deployment is very low and MPQUIC is not standardized yet, and therefore we found no single QUIC server supporting multipath in our list of targets (see Section~\ref{subsec:realworld}).

Previous work~\cite{lee2026quicstep,Govil2020mimiq} proposed to leverage QUIC connection migration to bypass \emph{on-path} censorship systems, i.e., blocking connections by detecting the SNI of a QUIC handshake. Our work differs in two key ways: we target \emph{on-server} geo-blocking rather than on-path SNI blocking, and we trigger migration only \emph{after the request headers have been accepted} rather than immediately after the handshake. Measuring both strategies toward a large number of destinations, we observe that existing on-path solutions, which migrate right after the handshake, fail to retrieve content for 34\% of requests that would be accepted when migrating after receiving the headers (see Section~\ref{sec:Evaluation}). In particular, we find that migrating right after the handshake does not enable to bypass geo-blocking on YouTube video streaming delivery.
This behavior follows from how videos are streamed. A video is delivered as a long sequence of chunk requests carried over the same connection, and the server checks each one as it arrives. Post-handshake migration moves the connection to the direct path once and leaves it there, so every chunk after the first is requested from the blocked address and refused. Bypassing a stream therefore requires migrating not once but once per request, which is what our post-header migration approach does.

We design, implement and evaluate \proxyname{}, named after the migratory bird, a solution to allow HTTP clients to benefit from optimized network path management thanks to connection migration without requiring any modification to the client itself.
\proxyname{} is a client-side proxy that translates HTTP/2 traffic from the local client into HTTP/3 over QUIC toward the remote server, enabling connection migration between request authentication and bulk transfer.
We show that \proxyname{} is compatible with major browsers and server implementations.
In a microbenchmark, \proxyname{} enables to receive 99\% of data over the direct path when downloading a 2MB file, while reducing the impact of a slower tunnel path on the overall download time.
In a real-world content restriction scenario, not all resources referenced by a web page support migration.
Sampling the Centralized Zone Data Service (CZDS)~\cite{czds}, we find that $5,684$ of the $33,561$ resolvable domains, that is 17\%, support QUIC. This is an upper bound on the domains any migration-based technique can be applied to, and it grows with HTTP/3 adoption. Over that set, half of the resources hosted on a server supporting migration can receive more than 75\% of their data over the direct path.
On YouTube, a real-world streaming service, \proxyname{} delivers 97.2\% of a geo-blocked video over the direct, blocked path with no impact on user experience.

In the rest of the paper, we first present the required knowledge to understand how the proxy works in Section~\ref{sec:background}.
The design and implementation of the proxy is presented in Section~\ref{sec:Architecture}.
Section~\ref{sec:Evaluation} presents how the proxy was evaluated both in a lab setting and in real-world scenarios.
Section~\ref{sec:discussion} then discusses how detectable \proxyname{} is, when it applies, and the consequences of the trust model.
Finally, we conclude in Section~\ref{sec:conclusion}.

\section{Background}
\label{sec:background}
We first review existing work on web censorship and its circumvention techniques.
Then, we present the state of the art on QUIC connection migration and multipath solutions, which form the basis of our more efficient mechanism for bypassing content restrictions.
We next introduce proxies, on which our solution builds, together with their performance implications.
Finally, we review related work on QUIC-based proxying and circumvention, positioning \proxyname{} with respect to these solutions.

\subsection{Web censorship and geo-blocking}

Censorship has been widely deployed across the world through various techniques to restrict access to specific content or services on the Internet~\cite{sundara2020censored,niaki2020iclab}.
There are two main layers at which censorship can be implemented: on-path and at the server.
On-path censorship aims to prevent clients from reaching the censored servers. It is usually the result of political decisions and is typically implemented directly by ISPs or through firewalls, ranging from enterprise to national scopes such as the Great Firewall of China.
Most on-path censorship circumvention techniques rely on obfuscating the traffic to reach an intermediate outside of the influence of the censor, such as a proxy or a VPN.
The connection is then tunneled through this point to reach the censored content.
Depending on the sophistication of the censorship, the obfuscation can range from simple IP tunneling to more complex techniques such as decoy routing~\cite{karlin2011decoy} or variation of it~\cite{sharma2020siegebreaker}.
QUICstep~\cite{lee2026quicstep} leverages QUIC connection migration to evade SNI-based on-path censors.

In this paper, we focus on destination geo-blocking, i.e., censorship implemented at the server, which typically relies on blocking access to specific content or services for specific connections.
Access to the content could be blocked because of distribution rights or to avoid dealing with data policies. An illustrative case is the many websites that blocked access to Europe when GDPR was introduced~\cite{403_forbidden}.
This is usually done by applying filtering rules based on the observed IP addresses of the client or on application-layer information such as HTTP headers.
The filtering can happen at the handshake, blocking the whole connection, or on each request, blocking only specific resources.
Circumvention techniques for server-side geo-blocking also rely on using intermediates to alter the perceived origin of the connection.


\subsection{QUIC migration and multipath}

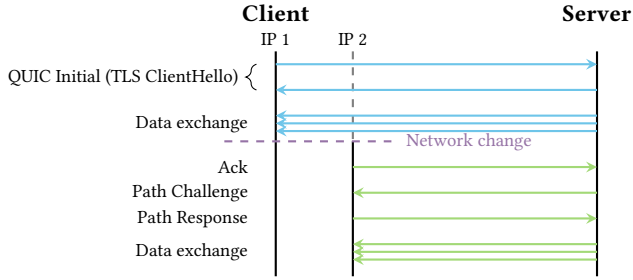
\begin{figure}[t]
\centering
    \begin{tikzpicture}[>=stealth, font=\small, scale=0.85]

  \definecolor{colblue}{HTML}{6EC5E9}
  \definecolor{colgreen}{HTML}{A3D977}
  \definecolor{colorange}{HTML}{D79B00}
  \definecolor{colpurple}{HTML}{9673A6}

  \newcommand{\cliA}{0}     
  \newcommand{\cliB}{1.2}   
  \newcommand{\srv}{5}      
  \newcommand{\h}{3.7}      

  \newcommand{\lbl}{-0.3}

  \node[font=\bfseries] at (\cliA, 0.4) {Client};
  \node[font=\footnotesize] at (\cliA, 0) {IP 1};
  \node[font=\footnotesize] at (\cliB, 0) {IP 2};
  \node[font=\bfseries] at (\srv, 0.4) {Server};

  \draw[thick] (\cliA, -0.2) -- (\cliA, -\h);
  \draw[thick, dashed, gray] (\cliB, -0.2) -- (\cliB, -1.6);
  \draw[thick] (\cliB, -1.6) -- (\cliB, -\h);
  \draw[thick] (\srv, -0.2) -- (\srv, -\h);


  \draw[->, colblue, thick] (\cliA, -0.4) -- (\srv, -0.4);

  \draw[->, colblue, thick] (\srv, -0.8) -- (\cliA, -0.8);

  \draw[decorate, decoration={brace, amplitude=4pt}]
    (\lbl, -0.8) -- (\lbl, -0.4);
  \node[anchor=east, font=\footnotesize] at ($(\lbl, -0.6) + (-0.15, 0)$) {QUIC Initial (TLS ClientHello)};

  \node[anchor=east, font=\footnotesize] at (\lbl, -1.32) {Data exchange};
  \draw[->, colblue, thick] (\srv, -1.2) -- (\cliA, -1.2);
  \draw[->, colblue, thick] (\srv, -1.32) -- (\cliA, -1.32);
  \draw[->, colblue, thick] (\srv, -1.44) -- (\cliA, -1.44);

  \draw[dashed, thick, colpurple] (-0.8, -1.6) -- (1.8, -1.6);
  \node[font=\footnotesize, anchor=west, colpurple] at (1.9, -1.6) {Network change};


  \node[anchor=east, font=\footnotesize] at (\lbl, -2.0) {Ack};
  \draw[->, colgreen, thick] (\cliB, -2.0) -- (\srv, -2.0);

  \node[anchor=east, font=\footnotesize] at (\lbl, -2.4) {Path Challenge};
  \draw[->, colgreen, thick] (\srv, -2.4) -- (\cliB, -2.4);

  \node[anchor=east, font=\footnotesize] at (\lbl, -2.8) {Path Response};
  \draw[->, colgreen, thick] (\cliB, -2.8) -- (\srv, -2.8);

  \node[anchor=east, font=\footnotesize] at (\lbl, -3.32) {Data exchange};
  \draw[->, colgreen, thick] (\srv, -3.2) -- (\cliB, -3.2);
  \draw[->, colgreen, thick] (\srv, -3.32) -- (\cliB, -3.32);
  \draw[->, colgreen, thick] (\srv, -3.44) -- (\cliB, -3.44);

\end{tikzpicture}
    \caption{Example of a QUIC connection when passive migration occurs due to a network change from the client. Each color represents a different network path. Upon network change, traffic leaves from the second interface (IF~2).}
    \label{fig:migration_example}
\end{figure}

MPTCP~\cite{rfc8684} was introduced to allow to join multiple TCP connections into a single session.
Because it relies on TCP headers that are in cleartext, middleboxes can easily interfere with connections, leading to very low deployment for web traffic~\cite{aschenbrenner2021single}.
QUIC~\cite{rfc9000} on the other hand experiences a rise in popularity for web traffic with all major actors supporting it~\cite{ruth2018first,Over9000}.
QUIC connections use Connection IDs (CIDs), which enable the connection to be identified independently of the IP address and port pairs enabling better privacy.
Therefore, the client can change their IP address during an ongoing connection without interrupting it.
This so-called connection migration process can be \textbf{actively} triggered by the client by sending a probing packet including a token on a new path using a different CID.
If the server receives the packet and allows the migration, it echoes back the token and probes along the new path to check the connectivity.
Once the new path is validated, both endpoints can start using it to send packets for the ongoing connection.
Because additional CIDs are exchanged encrypted, it is harder for middleboxes to identify or interfere with QUIC migrations.
The QUIC standard also allows \textbf{passive} migration that can happen when the perceived IP address of an endpoint changes, because of NAT rebinding for example.
In that case, the CIDs remain the same and the connection can just continue on the new path as long as the packets reach the server and the server supports migration.
Figure \ref{fig:migration_example} shows a QUIC connection where a network change occurs on the client side and a passive connection migration allows the server to validate the new path and continue the connection.

While not supported by all servers, QUIC connection migration is widely supported by most of the popular services~\cite{buchet2025analysis}.

\subsection{Proxy}
\label{sec:proxy}
Rather than implementing a custom QUIC client or modify a single browser, we choose to implement our solution as a proxy running on the client host. Section~\ref{sec:whyproxy} discusses the rationale behind this design choice.
Proxies are used as intermediates for HTTP connections. They can serve different roles based on their nature.
Local and public proxies can be explicitly asked by the client to forward their requests using the HTTP CONNECT method~\cite{rfc9110} to create an HTTP tunnel.
These explicit proxies can also intercept requests made by the client by forging TLS certificates to impersonate the server~\cite{mitmproxy}. This is only possible if the proxy has a certificate that is trusted by the client.

Ramesh et al.~\cite{Ramesh2024CalcuLatency} show that proxy use can be inferred by comparing the RTT of the application and the network-layer. On direct paths, both measurements reach the client, yielding a small RTT gap (typically <50 ms).
With a proxy, the network-layer probe stops at the proxy while the application-layer probe reaches the client, creating a larger gap (>50 ms).
In our setting, the proxy runs on the client host, so when using the direct path, both measurements terminate at the client and this heuristic does not apply.

The performance of HTTP connections using QUIC with a proxy have been studied by Liu et al.~\cite{liu2025performance}.
They find that, for small requests QUIC clients perform better than HTTP2 especially in networks with losses.
However, for larger requests or with increased latency, some QUIC implementations show degraded performance.


\subsection{Related work}

Several recent works have explored QUIC-based proxying and censorship circumvention techniques.
Kosek et al.~\cite{kosek2023secure} present a secured middlebox design prioritizing both security and performance, but not considering censorship scenarios.
They use the connection migration to decouple the connection between the client and the middlebox and the connection between the middlebox and the server.
The end-host is changing IP address frequently to prevent on-path tracking.
QuicCourier~\cite{huang2025quiccourier} leverages QUIC for packet manipulation to provide a covert channel.
By altering QUIC flows and adding a proxy layer, QuicCourier hides the nature of the traffic sent over the connection, making it indistinguishable from legitimate web traffic.

CoMPS~\cite{Wang2022splitting} relies on multiple paths and splits traffic on these paths using connection migration.
Using different network paths limits the information available to an on-path observer, making it harder to fingerprint the connection.
MIMIQ~\cite{Govil2020mimiq} and QUICstep~\cite{lee2026quicstep} both leverage QUIC connection migration against on-path adversaries.
MIMIQ preserves the privacy of an end-user against an on-path observer, but requires coordination in the client's network to handle the IP address changes and proper routing.
QUICstep focuses on evading SNI-based on-path censors, using a relay at the beginning of the connection before migrating to the direct path.
In contrast, our measurements show that on-server geo-blocking affects a significant number of destinations (see Section~\ref{sec:Evaluation}), and this is the primary focus of our work.
Because our proxy terminates HTTP traffic and has access to the response status code, it can time the migration based on application-layer feedback. It can also cover the on-path scenarios targeted by QUICstep, giving it a larger scope than those prior solutions.
Table~\ref{tab:related_work} highlights the differences between \proxyname{} and other solutions relying on QUIC connection migration for privacy and censorship circumvention.

\begin{table*}[t]
\centering
\caption{Comparison between \proxyname{}, CoMPS, MIMIQ and QUICstep. CoMPS and MIMIQ focus on fingerprinting protection with MIMIQ requiring ISP support. QUICstep only focuses on on-path censorship.}
\label{tab:related_work}
\begin{tabular}{lcccc}
\toprule
 & \textbf{On-Path censorship bypass} & \textbf{Server-side geo-blocking bypass} & \textbf{Fingerprinting protection} & \textbf{Requires ISP support} \\
\midrule
CoMPS & \xmarkr & \xmarkr & \cmark & \xmarkr \\
MIMIQ & \xmarkr & \xmarkr & \cmark & \cmark \\
QUICstep & \cmark & \xmarkr & \xmarkr & \xmarkr \\
\proxyname{} & \cmark & \cmark & \xmarkr & \xmarkr \\
\bottomrule
\end{tabular}
\end{table*}




\section{Design and architecture}
\label{sec:Architecture}
\begin{figure}[t]
    \centering
    \includegraphics[width=1\linewidth,trim=0 8 0 2, clip]{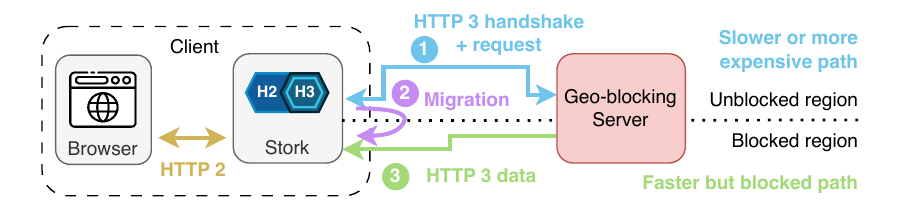}
    \caption{High-level architecture of \proxyname{}. A client-side proxy translates HTTP/2 traffic from the local client into HTTP/3 over QUIC toward the remote server, leveraging QUIC connection migration to separate handshake and request from bulk transfer.}
    \label{fig:architecture}
\end{figure}

This section outlines the architecture and key design decisions behind \proxyname.
The high-level architecture is illustrated in Figure~\ref{fig:architecture}.
Our design choices are made under the assumption that we run a local proxy, i.e., the proxy runs on the same host as the client application.
By colocating the client and the proxy on the same host, we ensure that the connection between the client and the server is not broken on any untrusted equipment.
The client connects in HTTP 1.1 using the CONNECT method to establish a tunnel to the proxy, then sends requests in HTTP/2.
This is a common setup for proxies, as it allows compatibility with most browsers and applications while keeping application changes minimal.
The HTTP/2 connection is intercepted by the proxy using a man-in-the-middle approach to terminate the TLS connection instead of the server.
We discuss the security implications of this choice in Section~\ref{sec:architecture:why_interception}.

The connection to the server is established by the proxy using HTTP/3 over QUIC, which enables leveraging QUIC's multiplexing and connection migration features for all traffic outside of the host.

\subsection{Threat model}
\label{sec:threat_model}

Our threat model is based on a client attempting to access a resource on a server that is subject to geo-blocking or on-path censorship.
We consider two types of adversaries. The first is a network-level censor, able to inspect and block any traffic on a direct path between the client and the server.
This censor, similar to the one described in QUICstep~\cite{lee2026quicstep} relies on SNI inspection and DNS filtering to prevent establishment of a connection to the server.
The second adversary is a server-side geo-blocker that applies filtering rules on a per-request basis, blocking specific resources based on the client's IP address or on application-layer information such as in HTTP headers.
In particular, the server-side geo-blocker only assesses the client's requests when they are received and is not aware of network-level changes such as connection migration.
This is the case for servers relying on reverse proxies or plugins to perform verification of the client's requests before forwarding them to the application server.
While a server or firewall could react to migration events and block the connection, this does not happen often in practice, as we observe with our real-world experiments targets in Section~\ref{subsec:realworld}. 

We assume that the client has access to an authorized path, outside of the control of the network-level censor and whose IP address is not blocked by the server-side geo-blocker.
The authorized path can be a VPN, a satellite link, or any other connection that allows the client to reach the server and retrieve the content without being blocked.
We assume that the authorized path has limited resources, either in terms of bandwidth for high-security solutions such as Tor \cite{tor} or in terms of cost for commercial solutions such as VPNs.
Studies showed that users in censored regions value speed over security for circumvention tools \cite{callanan2011leaping}, so the authorized path should be avoided when possible.
The client relies on a software proxy on the same host to manage paths used to reach the server.
The proxy uses a certificate trusted by the client and is assumed to be secure, i.e., it is not compromised by the adversaries and does not leak information about the client.
The traffic between the client and the proxy is assumed to be secure, i.e., it is not intercepted or modified by the adversaries.
Running the proxy on the same host limits the attack surface to malware on the host and ensures the proxy cannot be detected through cross-layer latency \cite{Ramesh2024CalcuLatency}.

\subsection{Internal design}

The proxy is implemented as a Rust program of about 3\,500 lines, largely relying on the quiche library~\cite{CloudflareQuiche} for QUIC and HTTP/3 support.
Each client connection is served by two threads, each running its own event loop: the client-facing thread owns the TCP connection and runs the TLS and HTTP/2 logic, while the server-facing thread owns one UDP socket per path and runs the QUIC and HTTP/3 logic.
Coordination between them uses message passing.

Clients do not send requests to the proxy directly, but first ask it to open a tunnel with the HTTP CONNECT method~\cite{rfc9110}.
Instead of forwarding the tunnelled bytes opaquely, \proxyname{} attempts to read incoming TLS ClientHello and generates on the fly a certificate for the requested server name, signed by a root certificate trusted by the client.
The server name is then used to open the QUIC connection to the actual server.

When the client opens an HTTP/2 stream, the proxy forwards the request headers to the HTTP/3 side, which opens a QUIC stream and keeps a mapping between the two stream identifiers; response data is looked up in that mapping and serialized back onto the corresponding HTTP/2 stream.
Not all frames have to be mapped between the two protocols.
Frames whose scope is limited to a single connection are handled locally, on the side where they are received. For instance, PING frames are answered on the connection that carried them, and each side maintains its own settings and header compression state.
Besides headers, data and stream termination, which are stream-scoped, only \texttt{GOAWAY} is forwarded, so that a shutdown on one side also terminates the other.

\subsection{Bypassing content restrictions on a single connection}

On the server-facing side, the proxy establishes a QUIC connection over an authorized path (e.g., a VPN or satellite link) to ensure the connection handshake is not blocked by content restrictions.
\proxyname{} supports two different migration timing: post-handshake migration and post-header migration.
In post-handshake migration mode, as done by QUICStep~\cite{lee2026quicstep}, the connection is migrated immediately after the handshake is completed and the requests are sent over a direct path that would otherwise be blocked.
Post-header migration is specific to \proxyname{}: the proxy waits for the headers of the HTTP requests to be sent on the authorized path to ensure the server accepts the request.
Once the server indicates acceptance via response headers, the proxy uses QUIC connection migration to move the ongoing connection to the direct path.
The bulk of the response is then transferred over a direct path as part of the same QUIC connection.

\smartparagraph{Active versus passive migration}
\proxyname{} uses passive connection migration, where the client's perceived IP address changes without actively notifying the server: the proxy simply switches the socket it sends from, mimicking a host moving between networks.
This is less invasive than active migration, which requires issuing a new connection ID for the new path and validating it with a \texttt{PATH\_CHALLENGE}.
On the other hand, the connection IDs are not rotated when the path changes, so an observer seeing both paths is able to link them. We discuss detectability further in Section~\ref{sec:discussion}.

\smartparagraph{Migration failure handling}
Even with passive migration, migration may fail for network-level reasons.
With anycast-based infrastructure, BGP routing may redirect packets to a different Point of Presence that does not know the connection state, triggering a \textit{Stateless Reset} packet to indicate that the connection is invalid (\cite{rfc9000}, Section~10.3).
Similarly, load-balancers using the 5-tuple for packet routing may hash the connection to a different backend server and send a reset.
In such cases, \proxyname{} detects the reset and returns to the authorized path.

When a server silently ignores the migration, or when the migrated path is blocked, no packet comes back at all.
\proxyname{} therefore arms a 1-second migration timeout, matching the initial Probe Timeout that RFC~9002 recommends when the RTT of a path is not yet known (\cite{rfc9002}, Appendix~A.2), and returns to the authorized path when it expires.
The (sub-)resource is then fetched entirely over that path.
Once a connection has reverted, \proxyname{} does not try to migrate it again, so a server refusing the path change only costs one timeout per connection.

\subsection{From single connections to complete web pages}
\label{sec:end2end}

\begin{figure*}[t]
    \centering
    \input{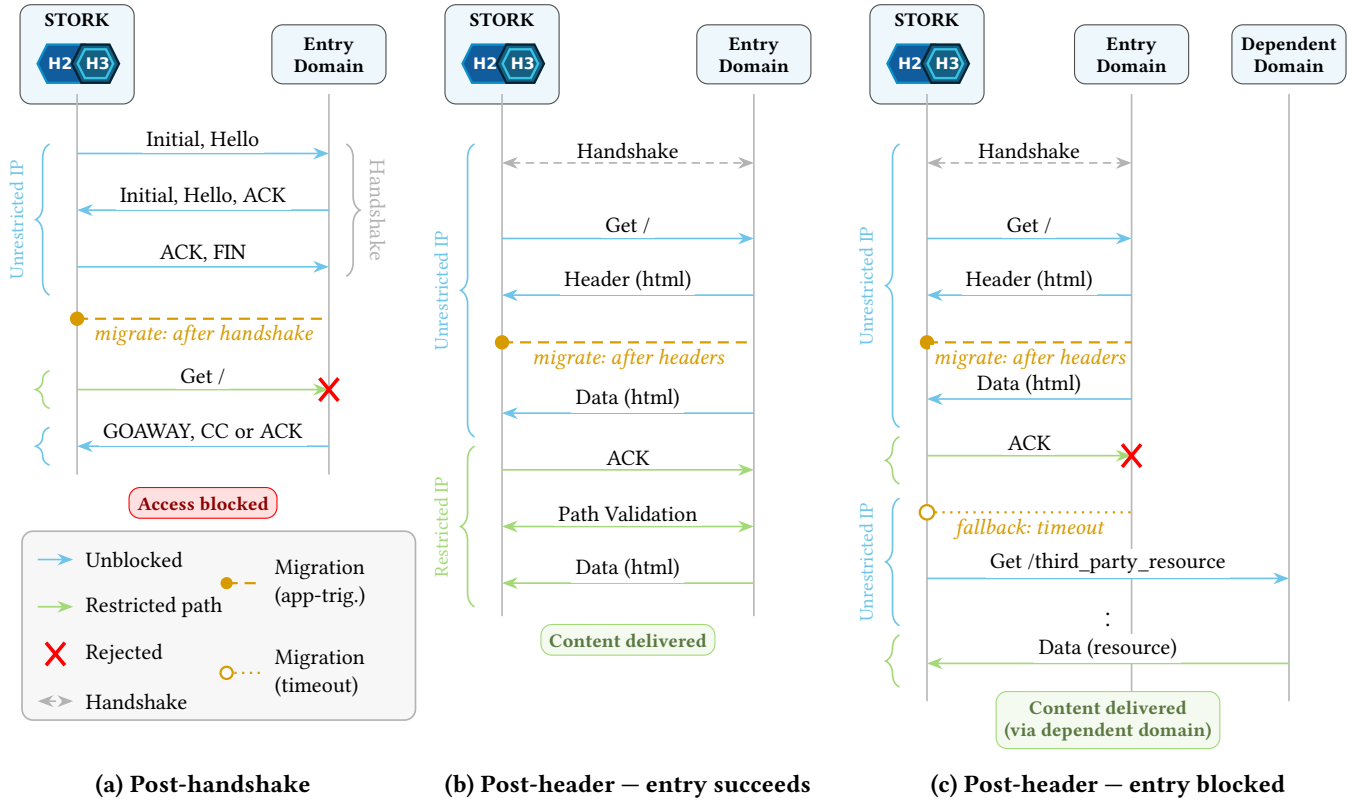}
    \caption{Migration patterns: (a) post-handshake migration~\cite{lee2026quicstep}, (b) \proxyname{} post-header migration when the entry domain succeeds, and (c) \proxyname{} post-header migration when the entry domain is blocked on the direct path.}
    \label{fig:post-handshake}
\end{figure*}

Figure~\ref{fig:post-handshake} (b) and (c) illustrate how \proxyname{} enables content restriction bypass for web services.
Most websites load resources from multiple domains, each with potentially different QUIC support and migration behavior.

When domain operators impose regional restrictions based on IP address, the server denies access with HTTP status codes like 403 or 451.
\proxyname{} first establishes the connection over the authorized path and sends the HTTP request on this path.
Upon receiving response headers indicating the server has accepted the request, \proxyname{} initiates migration to the blocked path.
If migration succeeds, the remaining data is delivered via the blocked path, effectively bypassing geo-blocking while reducing authorized-path usage.

To decide whether a domain speaks QUIC at all, \proxyname{} arms a handshake timeout, set to the same 1-second initial PTO used for the migration timeout.
If no QUIC connection can be established within that window, \proxyname{} concludes the domain cannot serve the resource over HTTP/3 and falls back to HTTP/2 over TCP on the server side, rather than letting the browser wait.
For domains that do support QUIC but not migration, \proxyname{} automatically uses the authorized path for the entire resource transfer.
\proxyname{} will attempt migration on the dependent domains (see Figure~\ref{fig:post-handshake} (c)).
By attempting migration for all supporting domains, \proxyname{} maximizes data delivered over the direct path while maintaining compatibility with heterogeneous server behaviors.

To avoid unnecessary migration, \proxyname{} uses heuristics to determine whether a migration would be beneficial, based on the HTTP headers of the response. Migration only occurs for responses with a 2xx status code, indicating the request was successful and the server accepted it. In case of a redirection, the migration does not occur directly, but only after the redirected request is accepted by the server.
In addition, migration is skipped when the response announces less than 20~KB of content, as it is then likely to be delivered before the migration completes; the threshold is based on the synthetic experiments of Section~\ref{subsec:synthetic} and applies to responses carrying a \texttt{Content-Length} header.
Responses with a \texttt{text/*} content type are never migrated, so the HTML document of a page is fetched over the authorized path and only its sub-resources are candidates for migration. This is because the HTML document is typically small and contains the references to all sub-resources, so migrating it would not save much time while risking a migration failure that would require a full re-fetch over the authorized path. We also observe servers often do not send a Content-Length for the main HTML page (likely generated on the fly), so the 20~KB threshold would not apply and the migration would be attempted, which is not desirable.

\subsection{Why a proxy?}
\label{sec:whyproxy}
We observe that most browsers do not implement client-triggered QUIC connection migration, even though it is part of the QUIC standard.
Moreover, most browsers default to HTTP/2 when initially connecting to a server~\cite{QUICeyeballs}, using HTTP/3 only when support is indicated either through DNS HTTPS records~\cite{rfc9460} or Alt-Srv headers~\cite{rfc7838}.
The proxy approach works around these browser limitations and additionally enables legacy software that only supports HTTP/2 to take advantage of QUIC features without requiring application changes.
The downside is that the client needs to trust the proxy, as it terminates the HTTP/2 connection and establishes a new HTTP/3 connection to the server.
This is why we recommend the use of the proxy directly on the client host.
We also plan to release \proxyname{} as an open-source project to allow users to verify its security.
Our proxy works with major browsers and is sufficient to demonstrate the benefits of QUIC connection migration in real-world scenarios. Future work could integrate this content restriction bypass, which builds on connection migration, directly into browsers or operating systems.

\subsection{Why interception?}
\label{sec:architecture:why_interception}
TLS interception is a controversial practice as it breaks the end-to-end encryption model of TLS.
However, in our use-case, the proxy is running on the same host as the client application. This is a common approach for applications monitoring web traffic such as antivirus~\cite{de2016killed,waked2018intercept}.
Durumeric et al.~\cite{durumeric2017security} find that 4 to 10\% of TLS connections received across multiple sets of servers had been intercepted, and formulate a series of recommendations to improve the security of client-side proxies.
Our proxy respects the latest TLS standards, including secure cipher suites and proper certificate validation in accordance with the recommendations of Durumeric et al.~\cite{durumeric2017security}.
Section~\ref{sec:discussion:trust} enumerates the attack surface this adds.

Beyond these security considerations, on-host interception is also a functional necessity: to trigger migration at the right moment, \proxyname{} needs access to information that an in-network proxy cannot obtain, since the latter only observes encrypted QUIC packets.
In particular, an in-network proxy cannot read the HTTP request nor see the response headers indicating whether the server accepted it, and could therefore neither time migration correctly nor fall back on the authorized path when the request is rejected.
Running the proxy on the client host instead gives \proxyname{} direct access to the HTTP status code, which underlies the path-selection logic described in Section~\ref{sec:end2end}.
A possible alternative that we leave as future work would be to infer request status from side-channel features such as packet sizes or inter-arrival times.
Prior work already demonstrates Machine Learning-based techniques for analyzing encrypted HTTP traffic~\cite{alwhbi2024encrypted} which could be adapted to determine the best moment to trigger migration without relying on header access.



\section{Evaluation}
\label{sec:Evaluation}
In this section, we present the evaluation of \proxyname{}.
The main features and the soundness of our design are verified in a controlled environment (Section~\ref{subsec:synthetic}).
The practical benefits and coverage are also evaluated through a set of real-world experiments performed from different locations around the world (Section~\ref{subsec:realworld}).

\subsection{Synthetic}
\label{subsec:synthetic}

We first conducted a series of experiments in a minimal controlled environment.
The objective of these experiments is to validate the ability of \proxyname{} to work with different QUIC servers and bypass geo-blocking.
While the system is not designed to have production-level performance, we also evaluate the overhead introduced by the proxy to ensure that it is reasonable.
We deployed a set of three VMs in a private cloud: an HTTP client using \proxyname{} as its proxy, a QUIC server and a VPN relay.
All VMs are running Ubuntu 24.04 using Intel Xeon Gold 6430 CPUs.
The client has 8 cores and 16GB of RAM allocated, the server has 2 cores and 8GB of RAM, and the relay has 1 core and 2GB of RAM.
The client uses two paths to the server: a direct link, and a virtual tunnel through the relay.
We used WireGuard~\cite{wireguard} to implement the tunnel; it is a lightweight VPN solution used by many providers, including the one in our real-world experiments.
We used the puppeteer JavaScript library~\cite{puppeteer} with headless Chromium~\cite{HeadlessChrome} to perform HTTP requests.

\smartparagraph{Server compatibility}
We start by validating the behavior and compatibility of \proxyname{} with different QUIC servers.
The proxy relies on the Cloudflare quiche library for HTTP/3, which also provides an example quiche server with migration enabled~\cite{CloudflareQuiche} that we use for the tests.
We also include the Picoquic server implementation, an open-source project supporting migration and known to be very mindful of the QUIC standards~\cite{PicoQuic}.
The last server used for validation is Nginx \cite{reese2008nginx}, a popular web server providing HTTP/3 support.

For each server, we verify that (i)~the proxy is able to establish a connection with the server, (ii)~the client is able to send a request through the proxy and receive the response, (iii)~the server is able to handle a passive connection migration triggered right after the handshake, and (iv)~the server is able to handle a passive connection migration triggered after the request headers were received by the proxy.

We also test whether the QUIC server can detect the migration and surface it to the application layer. This would allow the application to react to the new client IP and, for example, re-apply geo-blocking or adjust the content so that the migrated path remains blocked.
The results of these tests are summarized in Table~\ref{tab:server_compatibility}.
We were able to verify that all servers are working as expected and able to respond to all requests even when connection migration occurs.
For the notification of migration, Quiche connections keep a queue of path events that can be used by the application to detect and react to a migration.
Picoquic uses callback functions that can be called for specific events, allowing the application to choose whether to react to a migration.
At the time of writing, Nginx does not expose or log any migration-related events.
This means that applications using Nginx are not able to detect a migration and change their behavior upon a migration.

Our findings confirm that one might evade geo-blocking or geo-tailoring of content served by Nginx, while servers built on Cloudflare Quiche and Picoquic may leverage the migration notification to prevent evasion.
This could concern a significant portion of servers, since as of April 2026, Nginx is the most widely used web server, powering 32.7\% of websites~\cite{w3techs}.

\begin{table}[t]
\centering
\caption{Server compatibility.}
\label{tab:server_compatibility}
\begin{tabular}{lccc}
\toprule
 & \textbf{Quiche} & \textbf{Picoquic} & \textbf{Nginx} \\
\midrule
QUIC Handshake            & \cmark & \cmark & \cmark \\
HTTP/3 Request            & \cmark & \cmark & \cmark \\
Migration after handshake & \cmark & \cmark & \cmark \\
Migration after request   & \cmark & \cmark & \cmark \\
Notification of migration & \cmark & \cmark & \xmarkr \\
\bottomrule
\end{tabular}
\end{table}

\begin{figure}[t]
    \includegraphics[width=.95\linewidth, trim=0 0 0 10,clip]{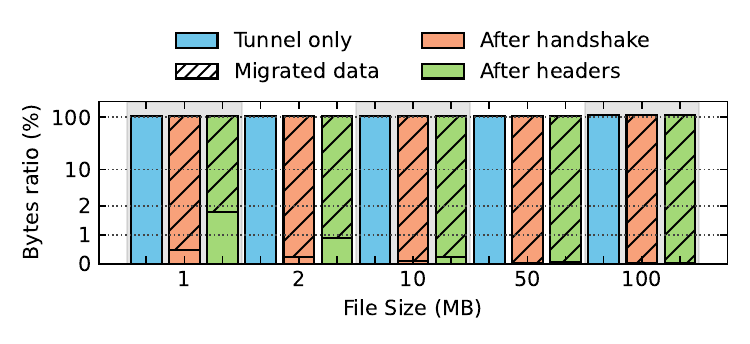}
    \caption{Ratio of file size sent over each path when not migrating, migrating right after handshake and migrating after headers. Symlog y-scale.}
    \label{fig:timing_difference}
\end{figure}

\smartparagraph{Migration timing}
To evaluate the impact on the volume of traffic, we request files of varying sizes under the three different settings supported: no migration, migration immediately after the handshake, and migration after the response headers are received.
Each experiment is performed 10 times. The mean number of bytes received on the primary interface and on the interface used for migration (hatched bars) is reported in Figure~\ref{fig:timing_difference}.

The total amount of bytes received when using migration, regardless of the timing, is almost the same as the regular request, meaning that there is a negligible overhead in terms of volume.
When migrating as soon as the connection is established, only the handshake packets use the primary interface, so only a few kilobytes.
Waiting for the headers to be received implies that the headers and the beginning of the response will also be sent over the relay interface.
Indeed, we wait for the reception of the response headers before performing the migration, to make sure the request is processed by the server.
The amount of bytes sent for the response should be roughly equal to the size of the congestion window of the server.
The initial congestion window for QUIC should be ten times the maximum datagram size, so around 12~KB~\cite{rfc9002}.
The window might also slightly increase during the handshake provided that there are no losses.
In our results, the total amount of data received on the VPN path ranges between 20 and 30 KB, no matter the total amount of data.
Even with 1~MB of data and waiting for the headers to be received, 98\% of the data is sent over the migrated path. For larger files, the volume sent over the primary path is negligible.

\begin{figure}[t]
\includegraphics[width=.95\linewidth]{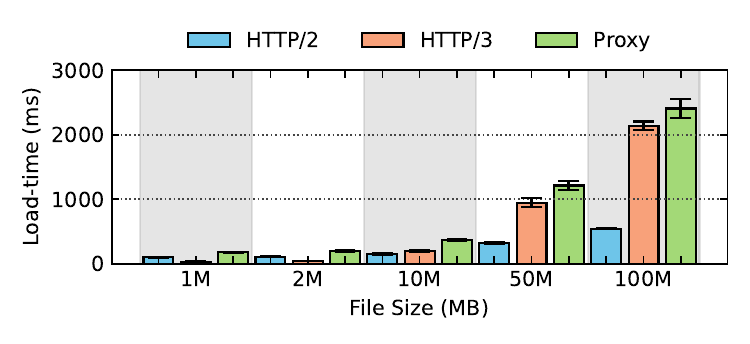}
\caption{Performance difference between direct HTTP/2, direct HTTP/3 and using the proxy for increasing transferred file sizes. Logarithmic scale for y-axis}
\label{fig:performance_difference}
\end{figure}

\smartparagraph{\proxyname{} latency overhead}
We then study the impact of the proxy on download time to ensure that the overhead introduced by the proxy is reasonable.
We measure the time to request files of different sizes from the server for different scenarios. (i) Direct HTTP/2 connection to the server. (ii) Direct HTTP/3 connection to the server. (iii) HTTP/2 to HTTP/3 translation through \proxyname{}.
The results are shown in Figure~\ref{fig:performance_difference}.
Each request is performed 10 times.

As expected, for requests smaller than 10 MB, the QUIC client is performing better than the HTTP/2 client.
For these requests, the overhead of using a proxy is relatively low.
For larger requests, the HTTP/3 client starts to get slower than HTTP/2.
This is in line with previous studies showing that some HTTP/3 implementations have worse performance than HTTP/2 for large transfers \cite{yu2021dissecting}.
Using the proxy shows the same trend: the overhead becomes more significant as the request size increases.
As the underlying QUIC connection slows down on larger transfers, this slowdown is also observed with the HTTP/2 to HTTP/3 proxy.
The overall performance remains in line with a direct QUIC connection.

\smartparagraph{Impact of Delay}
The last factor we study is the impact of the delay between the client and the server, as well as between the client and the relay.
As VPNs and relays are often used to access endpoints in different countries, sometimes far away from the client and/or the server, using a relay for the full connection might increase the latency significantly.
Using migration to switch from the VPN to the direct path after the request can therefore mitigate most of the added delay.

\begin{figure}[t]
    \centering
    \includegraphics[width=0.95\linewidth]{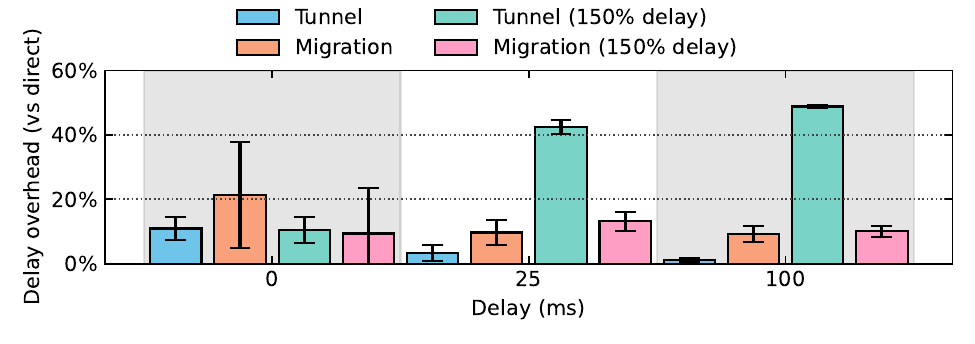}
    \caption{Impact of delay on the performance of the proxy under different conditions.}
    \label{fig:delay_impact}
\end{figure}

We measure the time to perform a request under different delay conditions on the client-server and client-relay links, and compare the results with the direct link.
Since all VMs are directly connected within the same cloud, the base delay is negligible, below 1ms.
The results are shown in Figure~\ref{fig:delay_impact}.
The additional delay is introduced using the \texttt{tc} command by adding a delay to the outbound traffic on both sides of each link.
We use a file size of 2 MB as it is the median size of a web page~\cite{httparchive}.

Requests using the tunnel take almost the same time as using the direct link, highlighting that the overhead of using Wireguard is very low.
We also conduct an experiment where we start the request through the tunnel and migrate the connection to the direct link after headers are received.
For this case, we use the same delay for both the direct link and the relay link.
The results show that the migration introduces a slight overhead corresponding to the round-trip time needed to complete the migration.
As we expect the delay to be higher when using a distant relay, we also measure the performance when the tunnel link has a delay that corresponds to 1.5 times the delay of the direct link.
The increased delay causes an overhead of about 50\% at 100ms of delay, while the load time with migration remains the same as with the same delay on both links, showing that the migration successfully avoids the added delay of the tunnel link.

\smartparagraph{Cloudflare WAF}
After establishing that the proxy is able to perform requests and migration with different servers with reasonable performance overhead, we now verify that it is able to evade geo-blocking.
To analyze the ability of \proxyname{} to bypass content filtering, we conduct experiments using Cloudflare's Web Application Firewall (WAF) \cite{cloudflareWAF}.
For this, we open our server, located in Europe to the public Internet and host a domain that we register with Cloudflare.
We use a client based in \asia{} with a VPN tunnel to a relay in Europe and test the ability of \proxyname{} to bypass different WAF rules using different migration timings.
We blocked access from our client IP address using different rules based on the country, IP prefix and IP address of the requests. We verify that we are able to access content on the server using \proxyname{} with both post-handshake and post-header migration in all cases.
This indicates that Cloudflare's WAF only verifies source IP addresses rules at the beginning of the connection and does not re-evaluate them after migration even when a request is sent from a different IP address.
Even when adding conditions on specific request headers, the filters on IP addresses are still not re-evaluated after migration.

\smartparagraph{Wordpress plugin evasion}
To further test the evasion, we use the IP2Location \cite{ip2location} plugin for Wordpress, a popular plugin to restrict access to websites based on IP addresses.
We deploy a Wordpress instance on top of the server, install the IP2Location plugin and configure it to block access to the website for the IP address of our client machine but allow access from the relay.
We verify if performing a migration from the tunnel to the direct link allows the client to successfully access the Wordpress instance while using only the direct link causes the request to be blocked.
We test the migration both directly after the handshake and waiting for the headers to be received.
Unlike with Cloudflare WAF, when migrating after the handshake, the request is blocked by the plugin, indicating that the check is performed for each request instead of once per connection.
As a consequence, solutions such as QUICstep~\cite{lee2026quicstep} that migrate right after the handshake are unable to evade this form of server-side geo-blocking, since the request still reaches the plugin from the direct (blocked) IP. Only our post-header migration, which first obtains acceptance on the authorized path, bypasses the plugin.

\subsection{Real-world experiments}
\label{subsec:realworld}

We conduct real-world experiments to validate both the geo-blocking bypass capability and performance of \proxyname{}.

\begin{figure}[t]
\centering
\includegraphics[width=0.8\linewidth]{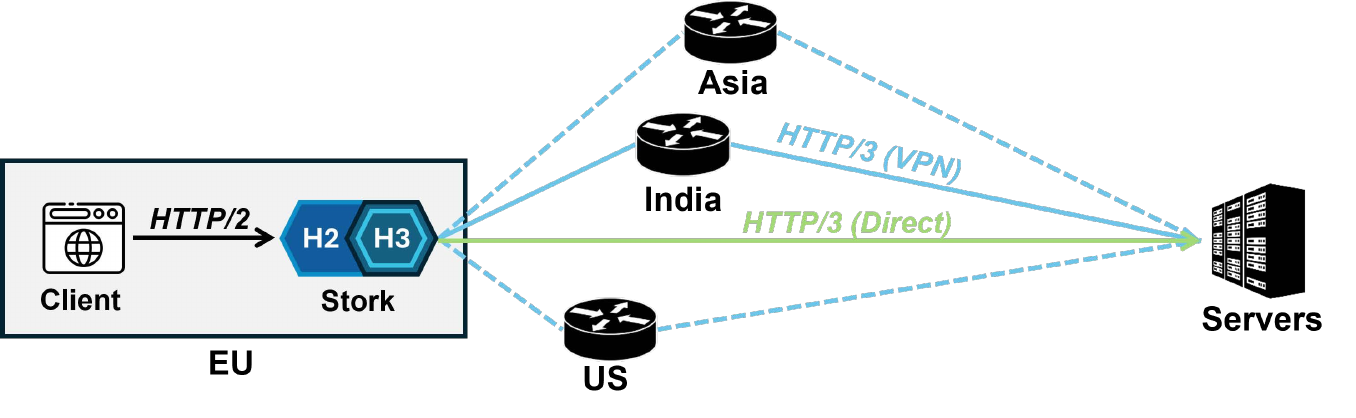}
\caption{Global testbed setup.}
\label{fig:global_testbed}
\end{figure}

\smartparagraph{Testbed}
Figure~\ref{fig:global_testbed} illustrates our testbed setup.
All experiments are run using a headless Chromium instance, which offers CLI access to the engine of the Chrome browser \cite{HeadlessChrome}.
This allows us to simulate real user browsing behavior and accurately measure page load times and data reception, including for sub-requests.
The client machine is located in Europe with a direct path to the Internet.
We used NordVPN to provide IP tunnels in \NA{}, India and \asia{}.


\smartparagraph{Targets}
\label{sec:Evaluation:list}
We randomly sample $100,000$ domains from the list of top level domains (.com, .net and .org) gathered from the Centralized Zone Data Service (CZDS)~\cite{czds}.
%
We first resolve domain names using massdns~\cite{massdns} with post-processing to resolve CNAME records.
This gave us $33,561$ unique IPs and domains with valid A records.
We probed domains using the Cloudflare \texttt{quiche} client~\cite{CloudflareQuiche} to detect QUIC support. We successfully completed a handshake with $5,684$ out of the $33,561$ servers.
%
To motivate our choice of QUIC instead of MPTCP, we also build an mptcp client and connect to all hosts responsive on the TCP port 80.
We only could establish an MPTCPv1 handshake with $330$ out of the $33,561$ servers, that is $0.9\%$.
We also propbe for multipath QUIC support on all QUIC targets (i.e. sending the \verb|initial_max_path_id| transport parameter in the handshake), and found no single server answering with a valid multipath QUIC handshake out of the 33k servers.

\smartparagraph{Geo-blocking domain identification}
We first identified geo-blocked domains in our target set to demonstrate the effectiveness of \proxyname{}. We followed the methodology proposed in \cite{length_comparison,403_forbidden} to classify domains based on their accessibility from different regions.
The initial step identifies suspicious domains by comparing the length of the main HTML body retrieved from different regions. If the body retrieved from one region is at least 30\% smaller than that retrieved from another, the domain is considered potentially geo-blocked~\cite{length_comparison}.
%
Differences in HTML body length alone are insufficient to confirm geo-blocking, as they may result from bot detection mechanisms. To address this, prior work applies TF-IDF (Term Frequency–Inverse Document Frequency) to represent HTML content as vectors and cluster domains with similar content patterns \cite{403_forbidden}.
Finally, each cluster is manually inspected to filter out false positives caused by bot detection, resulting in a refined list of geo-blocked domains. Following this methodology on our testbed, we identify a total of \textbf{111 geo-blocked domains}: 40 domains blocked in India, 26 blocked in EU but accessible in the US, 23 blocked in the US but accessible in EU, 14 blocked in Asia but accessible in EU, and 8 blocked in EU but accessible in Asia.

\begin{figure}[t]
\centering
\includegraphics[width=.95\linewidth,trim=0 4 0 4,clip]{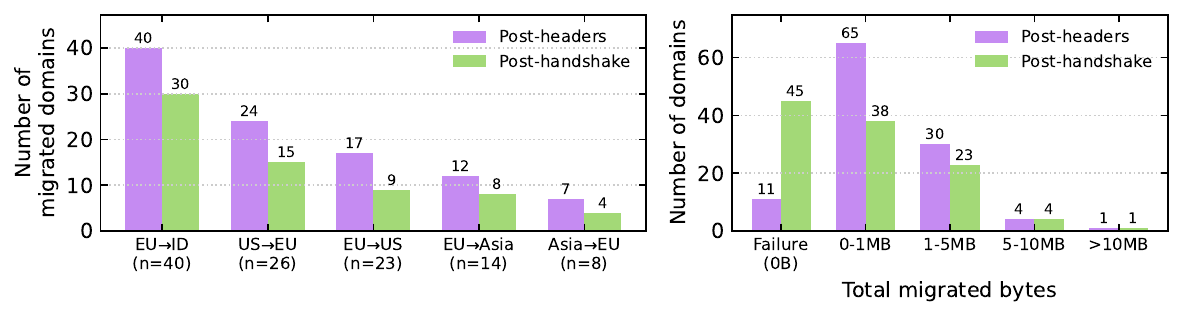}
\caption{Comparison of migration policies for geo-blocking bypassing where n is the total number of domains blocked in the region. The blocked region is on the right side of the arrow.}
\label{fig:migration_policy_comparison}
\end{figure}

\smartparagraph{Post-handshake versus post-header migration}
We evaluate the bypass capability of \proxyname{} across the 111 discovered geo-blocked domains using two strategies: post-handshake migration and post-header migration. Figure~\ref{fig:migration_policy_comparison} shows both the proportion of data received over the geo-blocked path and the number of successfully bypassed domains.

Across all geo-blocking regions, post-header migration consistently achieves a higher success rate, bypassing 100 out of 111 domains, compared to 66 for post-handshake.
This is because with post-header migrations the proxy can still retrieve content via dependent domains even if the entry domain fails, whereas post-handshake fails immediately in such cases. The 11 domains left do not support migration, even inside the same country.
Post-header therefore enables migrating more data over the geo-blocked path.

\smartparagraph{Hosts of geo-blocked content}
We further analyze the hosts of the geo-blocked content to understand for which providers \proxyname{} is most effective.
We find that out of the 111 geo-blocked domains, a large majority (93) are hosted by only two providers: Cloudflare (77) and Sucuri (16).
The remaining domains are hosted by a variety of providers without any appearing more than twice.
We report for the two main providers the number of domains for which \proxyname{} successfully bypasses the geo-blocking and with which migration strategy in Table~\ref{tab:cdn_bypass}.

Although the Cloudflare WAF can be bypassed with both migration strategies, 16 domains hosted by Cloudflare can only be accessed using post-header migration, indicating that they likely use additional server-side checks such as the Wordpress plugin discussed in Section~\ref{subsec:synthetic}.
Two domains hosted by Cloudflare could only be accessed using post-handshake migration. This is because these two domains have very small HTML pages (<2KB), which prevents the migration from ever being triggered.
The firewall of Sucuri seems to have a different behavior, blocking all attempted requests when using post-handshake migration, while allowing access to 13 domains with post-header migration.
There are 2 domains hosted by Cloudflare and 3 domains hosted by Sucuri for which \proxyname{} fails to access the content, regardless of the migration strategy used.
This could be due to additional server-side checks that are not bypassed by \proxyname{} or servers not supporting migration.

\begin{table}[t]
\centering
\caption{Geo-blocking bypass for domains hosted by Cloudflare and Sucuri}
\label{tab:cdn_bypass}
\begin{tabularx}{0.9\columnwidth}{Xrr@{\hspace{6pt}}@{\hspace{6pt}}rr}
\toprule
 & \multicolumn{1}{l}{\textbf{Cloudflare}} & \multicolumn{1}{l}{\textbf{Sucuri}} \\
\midrule
\multicolumn{1}{l|}{Migration after handshake} & \multicolumn{1}{c}{58} & \multicolumn{1}{c}{0}\\
\multicolumn{1}{l|}{Migration after headers}   & \multicolumn{1}{c}{73} & \multicolumn{1}{c}{13}\\
\multicolumn{1}{l|}{Both} & \multicolumn{1}{c}{56} & \multicolumn{1}{c}{0}\\
\multicolumn{1}{l|}{Failure} & \multicolumn{1}{c}{2} & \multicolumn{1}{c}{3}\\
\bottomrule
\end{tabularx}
\end{table}

\smartparagraph{YouTube Case Study}
YouTube is the workload that separates the two migration strategies most sharply.
Access restrictions are applied per video, but playback is not a single transfer: it is a long sequence of HTTP requests for successive media chunks, issued every few seconds over the lifetime of the connection, and the server authorizes each of them on arrival.
A strategy that changes path once as proposed by QUICStep~\cite{lee2026quicstep} cannot serve this pattern.
Post-handshake migration moves the connection to the direct path immediately and leaves it there, so from the second chunk onwards every request reaches the server from the blocked address and is refused; the video stops.
Streaming therefore requires the per-request migration described in Section~\ref{sec:Architecture}: for each chunk, \proxyname{} returns to the VPN path to issue the request and, once the server has accepted it, migrates back to the direct European path for the chunk body.
The alternation repeats for the entire duration of playback.

To collect geo-restricted videos, we use the YouTube Data API as follows:

\begin{enumerate}
\item For each evaluated country supported by NordVPN, we retrieve the 100 most popular videos.
\item We select videos that are accessible in the target country but unavailable in our own country according to the regional restriction information provided by the API.
\end{enumerate}

Using this method, we collect \textbf{37 geo-restricted videos} from 12 countries, each with a playback duration of at least 10 seconds. We access each video through a VPN endpoint in the corresponding country and migrate its chunk transfers to the direct European interface. Using the HTML5 media API~\cite{whatwg_html}, we measure the playback buffer, rebuffering events, and the proportion of bytes transferred through the VPN and direct paths.

\begin{figure}[t]
\centering
\includegraphics[width=.95\linewidth,trim=0 4 0 4,clip]{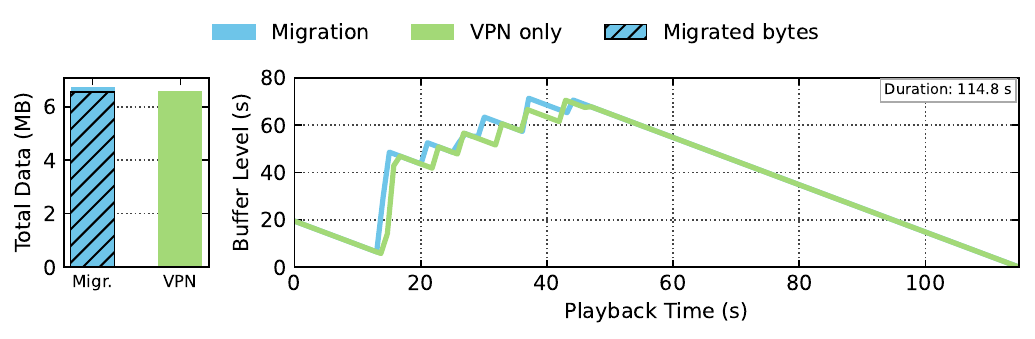}
\caption{YouTube case-study results comparing playback performance and migration efficiency.}
\label{fig:youtube_censorship_bypassing}
\end{figure}

We first examine the impact of \proxyname{} on these metrics for a representative video, accessible only from Canadian IP addresses.
Figure~\ref{fig:youtube_censorship_bypassing} compares \proxyname{} with VPN-only playback and reports both the playback buffer and path utilization.

The playback buffer remains above zero for both \proxyname{} and VPN-only playback, with no rebuffering events observed in either case.
The buffer levels are very similar. 
With \proxyname{} reaching higher buffer levels slightly faster than the VPN only approach, likely due to the lower latency of the direct path.
When the buffer level reaches the remaining playback duration, there is nothing left to load and the buffer decreases as the video plays.
Thus, post-header migration does not introduce additional playback interruptions.
At the same time, more than 90\% of the video data is transferred through the direct European path, substantially reducing VPN traffic without compromising playback performance.

\begin{figure}[t]
\centering
\includegraphics[width=.95\linewidth,trim=0 4 0 4,clip]{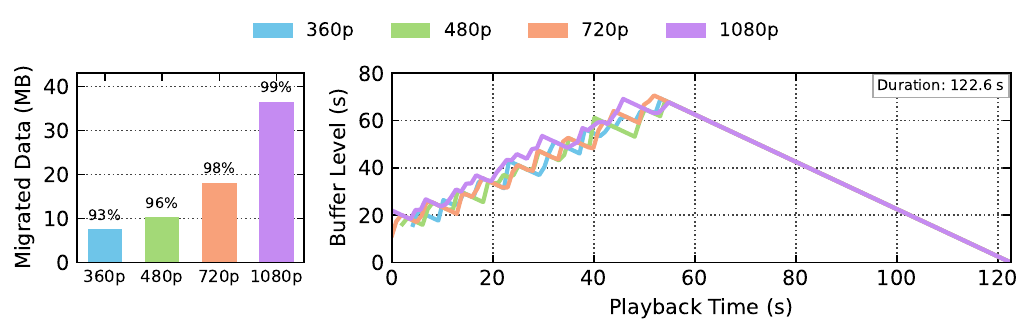}
\caption{Playback performance and migration efficiency across video resolutions.}
\label{fig:youtube_resolution}
\end{figure}

We next examine the impact of video resolution.
We select another video, accessible only from Japanese IP addresses, supporting resolutions from 360p to 1080p.
Figure~\ref{fig:youtube_resolution} shows the playback buffer and migration efficiency at each resolution.

No rebuffering event occurs at any evaluated resolution, and more than 90\% of the video data is transferred through the direct path in every case. 
Higher resolutions generate larger data volumes, increasing the absolute amount of traffic migrated away from the VPN while maintaining consistently high migration percentages.
This demonstrates the ability of \proxyname{} to provide good quality of service even for high-resolution streaming.

\begin{figure}[t]
\centering
\includegraphics[width=.95\linewidth,trim=0 4 0 4,clip]{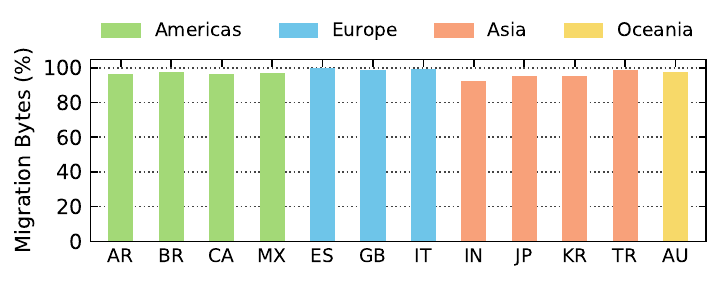}
\caption{Proportion of video data transferred through the direct path after migration, grouped by continent.}
\label{fig:youtube_migration_by_country}
\end{figure}

Finally, Figure~\ref{fig:youtube_migration_by_country} reports the proportion of bytes transferred through the direct path for all 37 videos, grouped by continent.

No rebuffering event is observed for any of the 37 videos, and more than 90\% of the data is transferred through the direct path in every case. Overall, \proxyname{} consistently bypasses YouTube geo-restrictions while preserving uninterrupted playback and carrying only a small fraction of the video traffic over the VPN.








\smartparagraph{Performance evaluation for VPN bypassing}

\begin{figure*}[t]
    \centering
    \begin{subfigure}[h]{0.48\linewidth}
        \centering
        \includegraphics[width=\linewidth]{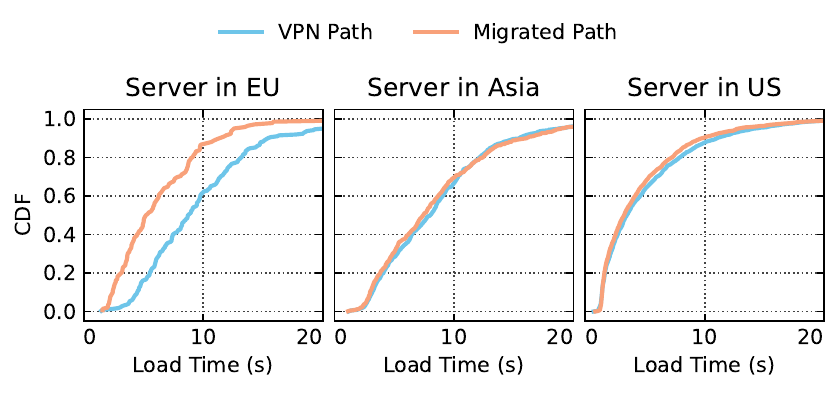}
        \caption{Asia to EU migration}
        \end{subfigure}
    \hfill
    \begin{subfigure}[h]{0.48\linewidth}
        \centering
        \includegraphics[width=\linewidth]{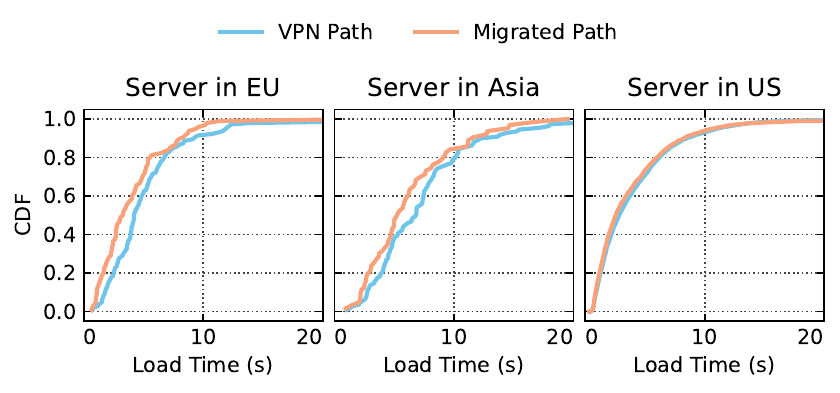}
        \caption{US to EU migration}
        \end{subfigure}
    \caption{CDF of page load times for VPN-only vs.\ VPN-to-direct migration}
    \label{fig:loadtime_comparisons}
\end{figure*}

We evaluate the impact of \proxyname{} on latency and bypass coverage across the 5K QUIC-enabled domains (not specifically geo-blocking ones) using different VPN endpoints. This enables an analysis among a more diverse set of targets.
For each domain, we first measure page load time using only the VPN path. We then repeat the experiment by establishing the connection over the VPN and migrating it to the direct path. DNS resolution is always performed via the VPN so that name resolution returns the server IPs a regular VPN user would obtain, rather than the one that would be returned from the client's local vantage point. For each setup, we run at least three measurements and report the median. Migration experiments are performed only if the VPN-only access succeeds. We include only domains where migration succeeds at least once, resulting in approximately $3,000$ domains for Asia to EU and $3,400$ domains for US to EU.

We use the ipinfo.io API~\cite{ipinfo} to determine the region of the server hosting the content. Figure~\ref{fig:loadtime_comparisons} then shows the CDF of page load times both per region-to-region migration (i.e., from the VPN region to the region of the client) and per region of the server.

When the VPN endpoint and the target server are geographically co-located, migration provides little latency benefit. In contrast, when they are located in different regions, migration improves performance by avoiding inefficient cross-country VPN routing. This effect is more pronounced in the (a) Asia-to-EU case, where the VPN endpoint is geographically distant from the EU; as a result, when the target server is located in Europe, migration yields significantly greater latency improvement.

\begin{table}[t]
\centering
\caption{Migration coverage distribution across domains} 
\label{tab:migration_distribution}
\begin{tabularx}{0.9\columnwidth}{Xrr@{\hspace{6pt}}@{\hspace{6pt}}rr}
\toprule
\% migrated content & \multicolumn{2}{c} {Domains (US$\rightarrow$EU)} & \multicolumn{2}{c}{Domains (Asia$\rightarrow$EU)} \\
\midrule
0--25\%   &  467 & (13.7\%)  &  479 & (15.9\%) \\
25--50\%  &  646 & (19.0\%) &  671 & (22.2\%) \\
50--75\%  &  306 & ( 9.0\%) &  282 & ( 9.3\%) \\
75--100\% & 1987 & (58.3\%) & 1588 & (52.6\%) \\
\midrule
Total     & 3406 & (100.0\%) & 3020 & (100.0\%) \\
\bottomrule
\end{tabularx}
\end{table}

Additionally, we highlight that in the Asia-to-EU case, more than half of the samples (52.6\%) achieved migration of over 75\% of the total data. Similarly, in the US-to-EU case, 58.3\% of domains fall into this category. This indicates that, when migration is allowed, more than half of the samples receive over 75\% of their data through the migrated path.

Three causes explain the other half of the samples, for which migration is less effective, and all of them are limits already discussed in this section. First, dependent domains that do not support QUIC fall back to TCP and are never eligible; Section~\ref{sec:Evaluation:list} shows that only 17\% of the domains complete a QUIC handshake. Second, resources below the 20~KB threshold are not migrated, as they complete before the path change takes effect. Finally, anycast-hosted resources answer the path change with a stateless reset, after which \proxyname{} returns to the authorized path for that connection.

Consequently, \proxyname{} reduces latency when the VPN and direct paths are geographically mismatched, while also significantly reducing VPN usage. On a complete page load, the benefit is however limited by the domains that do not support migration.


\section{Discussion}
\label{sec:discussion}
This section discusses the detectability of \proxyname{}, when \proxyname{} can be used, and the trust model.

\subsection{Detectability}
\label{sec:discussion:detectability}

We examine the detectability of \proxyname{} for the two adversaries defined by the threat model (Section~\ref{sec:threat_model}).

\smartparagraph{On-path observer}
Because \proxyname{} migrates passively, it does not rotate connection IDs when it changes path.
An observer that sees \emph{both} paths can therefore link the two flows on the CID alone.
In our threat model this observer does not exist: the authorized path is a tunnel whose payload the censor cannot read, so the CID used inside it is not visible on the direct path.
An observer that only sees the direct path has less to work with, but it can still notice short-header packets carrying a CID for which it never observed an Initial packet, and a stateful censor could flag them.
However, NAT rebindings and Wi-Fi to cellular handovers produce the same pattern, so a censor blocking on this signal would also block ordinary mobile users.
Rotating CIDs at the path change would remove the cross-path linkability. We however, measured less servers supporting active migration.

\smartparagraph{Server-side geo-blocker}
The stronger adversary is a server that re-evaluates the client IP when the path changes.
A server that re-evaluates it on \emph{every request}, as the WordPress plugin of Section~\ref{subsec:synthetic} does, can still be circumvented using the approach we apply to YouTube.
Nothing in QUIC prevents this, and our own results in Table~\ref{tab:server_compatibility} show that two of the three tested stacks already surface path events to the application: Quiche exposes a queue of path events and Picoquic invokes a callback.
An operator willing to act on them can re-apply its geo-blocking rule after migration and prevent any migration to a blocked address.
We however observe that operators do not do so in practice.
The Cloudflare WAF does not re-evaluate its IP rules after migration under any of the rules we configured (Section~\ref{subsec:synthetic}), and in the wild post-header migration retrieves the content of 100 of the 111 geo-blocked domains we identified.
Nginx, which powers 32.7\% of websites~\cite{w3techs}, does not expose migration to the application at all, so a geo-blocking plugin running on top of it has no way to react.

\smartparagraph{Transport fingerprint}
The upstream connection is opened by \proxyname{}, not by the browser, so the server observes the TLS and QUIC fingerprint of our HTTP/3 library rather than that of the client that will consume the content.
A server keeping track of \proxyname{}'s TLS fingerprint can notice the use of a proxy, as proposed by Frolov et. al. to detect circumvention tools~\cite{frolov2019use}.
This is a property of the proxy design and not of post-header migration. Integration into the browser or operating system would prevent it. As stated in Section~\ref{sec:proxy} timing-based detection~\cite{xue2024fingerprinting,Ramesh2024CalcuLatency} is not effective against a proxy residing on the same host as the client.

\subsection{When does \proxyname{} apply?}
\label{sec:discussion:scope}

One could expect Content Delivery Networks to defeat the migration approach, as anycast may route migrated packets to a Point of Presence that is unaware of the connection.
Geo-blocked content is, in our dataset, overwhelmingly CDN-hosted: 93 of the 111 geo-blocked domains sit behind only two providers, Cloudflare (77) and Sucuri (16).
Regional access control is a service that CDN providers offer, and it is applied at the application layer by a firewall or a plugin inspecting the request when it arrives.
Anycast also does not systematically break migration: 73 of the 77 Cloudflare-hosted domains are reached with post-header migration.
When re-routing does occur, the server sends a stateless reset that \proxyname{} detects, and the connection returns to the authorized path. The client therefore still receives the content, only without the benefit of migration.

The main limitation is QUIC availability: only $5,684$ of the $33,561$ resolvable domains we probed, that is 17\%, complete a QUIC handshake. The remaining domains are served over TCP, where the mechanism does not apply.
Resources below 20~KB are never migrated, as the transfer usually completes before the path change takes effect. This is why two Cloudflare-hosted domains with landing pages smaller than 2~KB are only reachable with post-handshake migration.
Finally, \proxyname{} assumes an authorized path exists at all; it makes that path cheaper to use, it does not replace the need for that path.


\subsection{Trust and attack surface}
\label{sec:discussion:trust}
Section~\ref{sec:architecture:why_interception} established that on-host TLS interception is common practice and that \proxyname{} follows the recommendations formulated for such proxies.
We describe here some of the consequences of interception.

First, the proxy installs a root certificate in the client's trust store.
This is the main risk introduced by our design, as any compromise of the proxy, or exfiltration of its signing key, allows an attacker to impersonate \emph{any} origin, including origins the user never routes through \proxyname{}.
The key is generated locally at install time and never leaves the machine, but a research prototype offers no protection beyond that.
Second, the proxy holds plaintext.
Every request and response transits in the proxy's memory in clear, therefore a compromised proxy can exfiltrate any content the user accesses through it. In our model, the proxy sits on the user's own machine, it is likely if the proxy is compromised locally, the whole machine is compromised and the attacker can read any content anyway.
Third, applications that pin certificates will refuse the forged chain.
Browsers accept locally installed roots for pinned origins~\cite{rfc7469}, but native applications that pin their own keys cannot be proxied at all and must be excluded from interception.
%
As noted in Section~\ref{sec:whyproxy}, a production design would move path selection into the browser or the operating system. Interception would then no longer be required, and none of the above would apply.

\subsection{Generality of our results}
\label{sec:discussion:generality}

Our real-world measurements use a single VPN provider to launch requests and detect geo-blocking from multiple locations.
A different provider would change the reputation of the exit addresses. If the ranges of that provider are themselves blocked, the request would already fail on the authorized path, before any migration takes place.
It could also change the tunnel overhead and, most visibly, the geographic position of the endpoint, which is the factor that determines how much latency migration recovers. Figure~\ref{fig:loadtime_comparisons} studies that dimension explicitly for two endpoints far apart.
A different provider would however not change the behavior \proxyname{} relies on, i.e. whether the server re-checks the source IP after a path change.



\section{Conclusion}
\label{sec:conclusion}

\proxyname{} leverages QUIC connection migration to bypass geo-blocking by initiating each connection on an authorized path and migrating it back to a normally blocked path once the server has accepted the request. In our synthetic experiments, up to 99\% of bytes for a 2~MB transfer are delivered over the released direct path. On the YouTube use case, where every chunk request is re-authorized by the server and migration must therefore be repeated for the entire duration of playback, our approach delivers more than 97\% of video chunks over the geo-blocked path with no noticeable degradation in playback quality.

The migration to the direct path may be detectable by advanced geo-blockers, potentially exposing the circumvention attempt, as discussed in Section~\ref{sec:discussion:detectability}.
We find however that most web servers do not notify the migrations and do not log them, hence plugins are unable to act on the migration and verify the location of the new client interface.
The advent of multi-pathing techniques in general with MPTCP, and MPQUIC in particular, opens new opportunities to generalise the approach presented in this paper. While we used a client-side proxy to demonstrate the feasibility of our approach, future work could explore integrating these capabilities directly into browsers, particulary to avoid TLS interception.




\section{Acknoledgments}
This research was partially supported by the European Union and Wallonia through the "Wallonia
2021-2027" program.

\bibliographystyle{plain}
\bibliography{biblio}

\appendix

\section{Ethical Considerations}
\label{sec:ethics}

We acknowledge that our work involves testing circumvention techniques in real-world censorship scenarios.
We did our best to not cause harm.
We did not target any particular government or organization's infrastructure, and our experiments were limited to publicly accessible services.
The scans involved only a few handshakes per destination, for 100k sampled domains in our initial QUIC detection pass.
We then fetched the home page for 5.6k domains for the geo-blocking detection pass, for each of the 7 scenario mentioned in the paper.
As each scan was done over multiple days, this does not pose any serious burden to the intermediate networks nor the targeted servers.

All experiments were conducted from our own infrastructure and the one of our VPN provider, NordVPN, which have lots of clients using their services to access censored content.
We did not involve recruiting participants from censored regions who could face legal repercussions.
Our proxy is designed as a research prototype to demonstrate feasibility rather than as a deployment-ready tool for mass circumvention.

Every endpoint was configured with a web server that clarified the purpose of our research experiments and included contact details for any inquiries or concerns.
We did not receive any complaints or requests to stop our experiments.
However, handshakes done through the VPN endpoint will not reveal the IP of our webserver. A deeper network analysis is required to identify our server IP address after migration, as discussed in Section~\ref{sec:discussion:detectability}.
This may have prevented our targets to consult our webserver.

While our work demonstrates techniques that could be used to bypass censorship, we recognize that the same mechanisms could potentially be misused.
Nevertheless, we believe the benefits of advancing censorship circumvention research and providing users with privacy-preserving access to information outweigh the risks.
Similar to prior work in this domain~\cite{fifield2015blocking,nasr2020massbrowser,sharma2020siegebreaker}, we operate under the principle that transparent research into circumvention techniques ultimately strengthens the ecosystem by informing both defenders of free expression and the broader research community.


\section{Open Science}
\label{sec:data}
The source code of \proxyname{}, is available on the following repository: {\url{https://anonymous.4open.science/r/Stork-2500/}}.
The instructions to run the experiments described in Section~\ref{sec:Evaluation} as well as the raw results obtained are provided within this repository: {\url{https://anonymous.4open.science/r/stork_experiments-3831}}.
We also provide a Docker environment that reproduces the controlled setup used in Section~\ref{subsec:synthetic} to facilitate replication.
The domain list used to gather the initial domains for the real-world experiments was fetched on the 27$^{th}$ of March 2026 from the CZDS website\footnote{\url{https://czds.icann.org/}}


\end{document}